\documentstyle[aps,prl]{revtex}
\begin{document}
\def\la{\langle}
\def\ra{\rangle}
\newcommand{\beq}{\begin{equation}}
\newcommand{\eeq}{\end{equation}}
\newcommand{\beqa}{\begin{eqnarray}}
\newcommand{\eeqa}{\end{eqnarray}}

\twocolumn
\begin{title}
{\Large \bf Transient interference of transmission and incidence}
\end{title}
\author{A. L. P\'erez Prieto$^1$, S. Brouard$^1$ and J. G. Muga$^2$} 
\address{$^1$ Departamento de F\'\i sica Fundamental II, 
Universidad de La Laguna, Spain}
\address{$^2$ Departamento de Qu\'\i mica-F\'\i sica, Universidad del 
Pa\'\i s Vasco, Apdo. 644, 48080 Bilbao, Spain}

\maketitle

\begin{abstract}
Due to a transient quantum interference
during a wavepacket collision with a potential barrier,  
a particular momentum, that depends on the potential parameters 
but is close to the initial average 
momentum, becomes suppressed. The hole left pushes the momentum
distribution outwards leading to 
a significant constructive enhancement
of lower and higher momenta. This is explained in the momentum complex-plane 
language in terms of a saddle point and two contiguous ``structural'' poles,
which are not associated with resonances but with  
incident and transmitted components
of the wavefunction.  
\end{abstract}

%

The traditional formulation of quantum scattering theory 
in terms of an ``$S$-matrix'' assumes that only the {\it results}
of the collision can be observed at ``asymptotic'' distances and times, 
but that the collision itself cannot be observed.  
This perspective is justified 
to analyze the products of standard collisions of 
atomic or molecular beams.
But the $S$-matrix approach is not enough to describe modern
experiments where the collision complex  
can be observed by means of femtosecond laser pulses or 
``spectroscopy of the transition state'' \cite{Zewail}. 
Also, in  quantum kinetic theory of gases accurate treatments must
abandon the ``completed collision''
approximation and use a full description, 
e.g. in terms of M\"oller wave operators as
in the Waldmann-Snider equation and its generalizations for moderately 
dense gases \cite{Snider}.     
In any case, it is important to understand the collision process itself   
to control or modify the products. This has      
motivated  a recent trend of theoretical and experimental
work to investigate the details of the collision itself,
and not only its asymptotics.
In particular, a quantum effect has been recently described by 
Brouard and Muga  \cite{BM98a,BM98b} in which the   
probability to find the particle with a momentum above a given value 
is larger, in the midst of the collision, than the quantity
allowed classically by energy conservation.
The effect  belongs to a group where the 
conservation of classical energy  
seems to be violated. Well known examples are the tunnel effect, 
or in general the non vanishing probability to find 
the particle in evanescent regions beyond the classical turning points.

This transient enhancement of the momentum tail  
may in principle be observed by collisions of ultracold atoms with a 
laser field  that can be turned off suddenly in the time scale of the atomic
motion \cite{BM98a}, and implies as a macroscopic 
consequence deviations from the Maxwellian 
velocity distribution \cite{BS}. We initiated the research of the present 
work looking for conditions that increase the effect and favor 
its observability. In so doing we have found an unexpected regime where 
the effect is much higher than in previously studied cases.
In this letter we shall describe such a regime and analyze
its physical origin, 
namely a 
transient interference between transmission and incidence
components of the wavepacket.
Let us first review briefly the main aspects of the 
classically forbiden increase of high-momenta.    
Brouard and Muga have studied several examples where the  
quantity  
$$
G^{\rm q}\left(p,t\right)\equiv\int_p^\infty\,\left\{|
\psi(p',t)|^2-|\psi\left(p',0\right)
|^2\right\}\, dp'
$$
takes on positive values for positive potentials
(the corresponding classical quantity is negative
or zero due to energy conservation) \cite{BM98a,BM98b}.
An important aspect of this effect is its {\it transient} character, 
$G^{\rm q}\le 0$ 
before and after the collision. 
The effect is also {\it generic} \cite{BM98a,BM98b}, 
because the stationary
components of the wavepacket have, in momentum representation,  
a tail due to the resolvent which is always present in the
Lippmann-Schwinger equation. This tail goes beyond the maximum value
allowed by the conservation of energy.
  
For a Gaussian wavepacket colliding with an infinite wall,
maximum values of
$G^{\rm q}_{max}(p,t)\simeq0.05$ have been reported \cite{BM98a}.
Also a ``delta'' potential was used \cite{BM98b} to analyze the influence 
of the opacity of the barrier.  
For the cases examined, an increase
of $G^{\rm q}$ with the opacity was observed up to a saturation 
level where the infinite wall results were recovered \cite{BM98b}. This 
suggested that the observability of the  
effect would improve with strongly opaque conditions.
 In a complementary study we have systematically 
varied the spatial variance of the wavepacket,
$\delta_x$, and the height of a square barrier, $V_0$, for  
a fixed average initial momentum $p_c$. We have found,  
contrary to previous expectations,
that the maximum effect corresponds to energies well above the barrier
and to large values of $\delta_x$.
In this regime the barrier is not at all opaque and
essentially the full wave is finally
transmitted. Moreover, $G^{\rm q}_{max}$ is as
large as $0.27$.
 
The numerical effort to perform these calculations   
by ordinary propagation methods (such as the split operator method) 
is rather heavy, since large values of $\delta_x$ and the need to discern 
fine details of the momentum
distribution require an extense and dense grid.         
In fact for very large values of 
$\delta_x$ this numerical route has to be eventually abandoned. 
But even if one gets numerical results with enough computer power,
they will not provide any explanation
of the unexpectedly high $G^{\rm q}$ values.   
Fortunately these two  difficulties can be overcome with an approximate
analytical solution. Here we shall sketch its obtention, 
a more detailed account will be given elsewhere.  
First the momentum representation of the wavefunction is expressed
using the basis of stationary eigenstates of $H$, $|p^{'+}\ra$, corresponding 
to incident momentum $p'$,  and energy $E'=p'^2/(2m)$, 
\beq\label{sdt}
\psi(p,t)=\int_{-\infty}^{+\infty}\,\la p|p^{'+}\ra e^{-iE't/\hbar}
\la p^{'+}|\psi(t=0)\ra \,dp'.
\eeq
If the initial state at time $t=0$ does not overlap with the 
potential and has negligible negative momentum components we can write 
\beq\label{asdt}
\psi(p,t)=\int_{0}^{+\infty}\,\la p|p^{'+}\ra e^{-iE't/\hbar}
\la p'|\psi(t=0)\ra\,dp'.
\eeq
To facilitate the treatment of the integral in the 
$p'$-complex plane we may now extend the lower limit  
to $-\infty$ using the analytical continuation 
of $\la p|p^{'+}\ra$, $p'>0$, over $p'<0$ (and later over 
the whole complex plane).   
          
For a square barrier of height $V_0$ and width $d$, centered at
the coordinate origin, the delta-normalized
stationary wavefunction with incident momentum $p'$ is 
\beq\label{xp'}
\la x|{p'}^+\ra=\frac{1}{h^{1/2}}\cases{
Ie^{ik'x}+Re^{-ik'x},& $\,\,\,  x<-d/2$\cr
Ce^{ik''x}+De^{-ik''x},&$\,\,\, -d/2<x<d/2$\cr
Te^{ik'x},&$\,\,\,x>d/2$,\cr}
\eeq
where $I=1$, 
$k'=p'/\hbar$ and $k''={\sqrt{p^{'2}-2mV_0}}/{\hbar}$.
The coefficientes $R,C,D$ and $T$ are determined
by continuity of the wavefunction and its derivative.   
The momentum representation $\la p|p'{^+}\ra$
will correspondingly have five terms. 
The terms with  $I$, $R$ and $T$ have poles in the $p'$-complex momentum
plane at 
\beqa\label{zabf}
&&p'_I=p+i 0\nonumber\\
&&p'_R=-p-i 0\nonumber\\
&&p'_T=p-i 0,
\eeqa
while the terms with $C$ and $D$
do not have these {\it structural poles} \cite{WS}, which are not related
to resonances or to the potential profile.     
The four functions $R,C,D$ and $T$ present an infinite series 
of {\it resonance and anti-resonance poles} in the 
third and fourth quadrants
due to the zeros of a common denominator
\beq
\Omega(p')=\cos\left(k''d\right)-
\frac{i}{2}\left(\frac{k''}{k'}+\frac{k'}
{k''}\right)\sin\left(k''d\right).
\eeq
(In particular $T(p')=\exp(-ik'd)/\Omega(p')$.)
The conditions examined in this work correspond however 
to ``direct scattering'',
where these resonance singularities do not play any
significant role.  

The initial state is taken as a minimum-uncertainty-product
Gaussian centered at the position
$-\alpha\delta_x$,  $\alpha>0$, with average momentum $p_c$, 
\beq\label{eig}
\la p'|\psi(t=0)\ra=\left(\frac{2\delta_x}{\pi\hbar^2}\right)^{1/4}
\,e^{-\frac{\delta_x(p'-p_c)^2}{\hbar^2}+
\frac{ip'\alpha\delta_x}{\hbar}}.
\eeq
This expression and the momentum representation of (\ref{xp'}) are 
inserted in (\ref{asdt}) to obtain five integrals. 
The full treatment of the resulting integrals follows closely ref. \cite{BM96}.
The integrals with $C$ and $D$
may be evaluated with the steepest descent  method 
for large values of $\delta_x$. 
The steepest descent path (SDP) is a straight line with slope    
$-t\hbar/(2m\delta_x)$, with a saddle point close to $p_c$ 
in the midst of the collision.
We shall always assume that the slope is small   
so that when the integration contour 
is deformed along this path it  ``cuts'' the
resonance poles of the fourth quadrant 
far from the real axis, and their residues can be neglected 
(``direct scattering'' conditions).   

Because of the interference between the saddle and the structural
poles, the 
simple steepest descent treatment valid for $C$ and $D$  
is not valid for the other terms. A uniform expression  
for a smooth treatment of the pole crossing of the SDP
is provided by the  
$w$-function, $w(z)=e^{-z^2}\,erfc(-iz)$, which may also be defined
by its integral expression \cite{Abram} 
\beq\label{wz}
w(z)=\frac{1}{i\pi}\int_{\Gamma_-} du\,\frac{e^{-u^2}}{u-z}, 
\eeq
where $\Gamma_-$ goes from $-\infty$ to $\infty$ passing below the pole. 
Since we are interested in wavepackets with energy well
above the barrier maximum
the ``reflection term'' with $R$ may be neglected.
The remaining contribution is   
\beq\label{iabf}
\psi_{IT}=ih^{-1/2}\hbar\tau\int_{-\infty}^\infty \,\left[g_I (p')+
g_T (p')\right]
e^{\phi (p')}dp',
\eeq
where 
\beqa
\tau&=&\frac{1}
{\sqrt{2\pi\hbar}}\left(\frac{2\delta_x}{\pi\hbar^2}\right)^{1/4}
\nonumber\\
g_I (p')&=&\frac{e^{ipd/2\hbar}}{p-p'+i 0}
\nonumber\\
g_T (p')&=&\frac{-T(p')\,{\rm exp}\left[i\left(2p'-p\right)d/2\hbar\right]}
{(p-p'-i 0)},
\nonumber
\eeqa
and
\beq\label{phi}
\phi\left(p'\right)=\frac{-i{p'}^2t}{2m\hbar}-\frac{\delta_x\left(p'-
p_c\right)^2}{\hbar^2}+\frac{ip'\left(\alpha\delta_x-d/2\right)}{\hbar}.
\eeq
The functions $g_I(p')$ and $g_T(p')$ present structural poles at $p'_I$
and $p'_T$; in addition $g_T(p')$ has resonance and anti-resonance poles.

The SDP is a straight line 
with the same negative slope as before, passing through
the saddle point,  
\beqa\label{saddle}
s&=&\frac{m}{4m^2\delta_x^2+t^2\hbar^2}\left\{4mp_c\delta_x^2+
\left(\alpha\delta_x-d/2\right)\hbar^2 t\right.\nonumber\\
&+&\left.i2\hbar\left[m\delta_x\left(\alpha\delta_x-d/2\right)-p_c
\delta_xt\right]\right\}.
\eeqa
To integrate (\ref{iabf}), the contour is deformed 
to the SDP passing over the saddle. 
The same reasons to neglect the residues from the resonance poles 
in the $C$ and $D$ terms are now applicable.   
To introduce the $w$-functions, the integrand must be put in the 
form (\ref{wz}). We complete the square in  
(\ref{phi}) and use the change of variable 
\beq
u=\frac{p'-s}{f}\,,\;\;\;\;\;\;\;
f=\left(\frac{\delta_x}{\hbar^2}+i\frac{t}{2m\hbar}\right)^{-1/2}
\eeq
to obtain 
\beqa
\nonumber
\left<p|\psi(t)\right>&\simeq& if\tau h^{-1/2}\hbar\,
e^{-\left(\delta_x p_c^2/\hbar^2\right)+\eta^2}
\\
&\times&
\int_{-\infty}^{\infty}\left[g_I\left(u\right)+
g_T\left(u\right)\right]e^{-u^2}\,du,
\nonumber
\eeqa
where $g(u)\equiv g(p'(u))$, and  
$$
\eta=\left(\frac{2p_c\delta_x}{\hbar^2}+i\frac{(\alpha\delta_x-d/2)}
{\hbar}\right)\left[4\left(\frac{\delta_x}{\hbar^2}+
i\frac{t}{2m\hbar}\right)\right]^{-1/2}.
$$
We may retain the main contribution from $g_I$  
from its behaviour near the pole
by approximating $g_{I}(u)\approx {\cal R}_{I}/(u-u_{I})$,     
where ${\cal R}_I$ is residue of 
$g_I(u)$ at the point $u=u_I=(p'_{I}-l)/f$, and similarly for $g_F$.   
For an approximate expression of $\la p|\psi\ra$, and considering that 
the wave is much more extended in space than the barrier we may neglect 
the contribution from $C$ and $D$ and retain only the incidence and
transmission terms,   
\beqa
\la p|\psi(t)\ra&\simeq& h^{-{1/2}} \pi
\tau\hbar\,e^{-\left(\delta_x p_c^2/\hbar^2\right)
+\eta^2}e^{ipd/2\hbar}
\nonumber\\
&\times&\left[w(u_I) + T(p) w(-u_T)\right]\equiv \psi_{IT}^0(p,t).
\label{psiIT}
\eeqa
A more precise expression including a term $\psi_{RCD}$ and corrections 
to the zeroth order $\psi_{IT}^0$ is given elsewhere and allows   
to obtain the wave function and $G^q_{max}$ accurately for
large values of $\delta_x$ with small 
computational effort.
However (\ref{psiIT}) captures the essential,   
and provides a simple, explanatory picture of the phenomenon
we want to discuss. 

Fig. 1 shows the distribution of momenta $|\la p|\psi(t)\ra|^2$
for different instants of time, from the 
initial one to a time after the collision has been completed, 
passing through the instant where $G^{\rm q}=0.27$ is maximum.
In all figures the numerical values correspond to collisions of ultracold
Rubidium atoms with an effective laser potential.    
The observed behaviour does not have a classical explanation.
Note that the wavepacket is considerably broader than the barrier.
A classical ensemble of particles with the same Gaussian phase-space 
(Wigner) distribution as (\ref{eig}) would only be slightly
deformed due to the small fraction of particles 
located on the barrier at a given time, and would keep   
the maximum at the average momentum  $p_c$.
Moreover, there could not be any spectacular acceleration 
or deceleration as the one seen in the two peaks of the
quantum distribution.
We shall see that the zero of the quantum momentum distribution,
which forbids in this case the initially dominant momentum
$p_c$,   
is due to a destructive interference whereas the two new peaks  
correspond to momentum regions of constructive interference. 

In Fig. 2 the {\it Argand diagrams} of the 
two terms are represented, namely the imaginary versus
the real parts obtained by varying $p$ at equal intervals. 
Each lobule corresponds to one of the terms.
The ``motion'' as $p$ increases begins close to the origin, downwards in both 
diagrams. The left peak of the momentum distribution corresponds to the  
zone where the two moduli increase together and are approximately 
in phase. After the descending motion there is a fast, aproximately
circular motion 
where the phases become opposed (destructive interference).
Finally, the two curves meet again {\it in phase}   
in the upper part of the lobules,
this momentum interval corresponds to the right peak
of the momentum distribution.
The described behaviour is essentially due to the two $w$-functions 
$w(u_I)$ y $-w(-u_T)$, as shown in Fig. 3, where  
the two Argand diagrams of the two $w$-functions 
and of the factors that multiply
them are represented between the momenta of the two 
maxima. Clearly the efect of the factors, whose phases remain
essentially constant 
around $\pi$, is simply to invert the two  
lobules of the $w$s.
The fast motion of the $w$-functions is due to the pass of the two
contiguous structural poles 
$u_I=(p+i0)/f$ and  $u_T=(p-i0)/f$ near the saddle point at $u=0$.
A sweep from smaller to larger values of $p$ moves 
the couple of poles along the real $p'$ axis from left to right, while,  
for fixed $t$, the saddle point and the steepest descent path 
do not depend of $p$. Since $u_I\approx u_T$ we can write, 
using the relation between $w$-functions of argument of opposite sign,
see (\ref{wz}), 
\beq
w(u_I)=e^{-u_I^2}-w(-u_T).
\eeq
During the collision, the saddle point is  
very close to the real axis of the $p'$-plane, only slightly below in  
Fig. 4, and    
the slope of the SDP  is very small.
This means that 
the difference between the two $w$-contributions 
is essentially a {\em real} exponential, 
which implies a ``simultaneous motion'', with equal maginary parts, 
along the two lobules of the Argand diagram. 
The rapid variation of the phases of the $w$s when passing close to the 
saddle point follows from its integral expression (\ref{wz}).
When $u_I$ passes close to $u=0$ and close to the 
real axis of the $u$-plane,  
the denominator is essentially real and changes its sign quickly, 
so there is a rapid change by $\pi$ in the phases of $w(u_I)$ and 
$-w(-u_T)$. 
The phase oposition alone does not explain 
however why the interference is totally destructive.  
It is also necessary to have equal moduli of the two incidence and
transmission terms of 
(\ref{psiIT}) for an exact cancellation. 
Actually the equality is obtained only transitorily, since 
before and after the collision only one lobule remains, the one for incidence 
before the collision, and the one for transmission after the collision.   
Along the collision the incident lobule decreases and  
the transmission one grows, until they equilibrate  
and give a perfect cancellation 
and the two constructive interference zones  
of Fig. 1.

By changing the barrier height the fases of the factors that multiply the  
$w$s change, the lobules rotate with respect to each other,
and one of the two in-phase regions grows while the other 
diminishes, so that the two peaks of the momentum distribution
become asymetric, see Figures 
5 and 6, where the momentum distributions and the 
corresponding lobules of the Argand diagrams are shown,
compare also with Fig. 2.      
Note that these factors do not depend on time and therefore the 
angle between the lobules remains constant 
throughout the collision. This means that the positions of the 
maxima and minima do not change significantly for a given collision.    

An important point is that the interference effect described
does not depend critically on 
the square barrier potential, and we have observed it in particular for a 
Gaussian barrier. Note that the arguments leading to
Eq. (\ref{psiIT}) are in fact 
of general validity and independent of the potential shape, with $-d/2$ and
$d/2$ being points where the potential may be 
assumed to be essentially zero, and $T$ being the corresponding transmission
amplitude.  
The possibility to observe this effect with ultracold atoms rests on the 
ability to prepare appropriate initial states. Turning off the laser 
potential during the collision will leave a two peaked momentum
distribution that implies 
at later times a visible spatial separation between two wave components,
one faster than the other.

We thank A. Steinberg for many useful discussions, and acknowledge 
support by Ministerio de Educaci\'on y Ciencia (PB97-1482).


{\bf Figure captions}

\small

{FIG. 1.} $\left|\left<p|\psi(t)\right>\right|^2$ for differents values
of $t$:
$t=0$ (dotted-dashed line);  
$t=2.333\, t_u$ (solid line); $t=2.731\, t_u$ (dashed line); and 
$t=3.233\, t_u$ (dotted line).
$m= 1.558023 m_u$, 
$V_0=102.5\, e_u$, $d=2.5\, l_u$, $-\alpha\delta_x=-50\, l_u$, 
$\delta_x=107.99\, l_u^2$, with an average 
momentum $p_c=28.48\, p_u$ well above   
the classical threshold $(2mV_0)^{1/2}=17.87\, p_u$.
The units are scaled for numerical convenience in the 
computations as $e_u=10^{-13}$ a.u. of energy, $p_u=10^{-4}$ a.u.
of momentum, $l_u=2\times10^{6}$ a.u. of lenght, $m_u=10^{5}$ a.u. of mass,
and $t_u=2\times10^{15}$ a.u. of time.      

{FIG. 2.} Imaginary versus real parts of the incident contribution to
$\psi_{IT}^0(p,t)$ (empty circles), and of the transmision contribution
(filled circles), for $t=2.731\, t_u$ and different values of
$p$ equally spaced between $p=28\, p_u$ and $p=29\, p_u$.
Other parameters as in Fig. 1.

{FIG. 3.} Imaginary versus real parts of $\omega(u_I)$ (empty circles)
and $-\omega(-u_T)$ (filled circles) for $t=2.731\, t_u$ and different values
of $p$ equally spaced between the two peaks of Fig. 1,
see the text. The prefactors corresponding to $\omega(u_I)$
and $-\omega(-u_T)$ in $\psi_{IT}^0(p,t)$ are also shown for
the same momentum interval,
solid and dashed lines respectively. Other parameters as in Fig. 1.

{FIG. 4.} Integration contour in the complex $p'$-plane when the SDP crosses
the pair of structural poles $p'_I$ and $p'_T$. Also shown the
structural pole  $p'_R$, the saddle
point of equation (\ref{saddle}), and the incident average momentum $p_c$.

{FIG. 5.} $\left|\left<p|\psi(t)\right>\right|^2$ as a
function of $p$, for
two different values of $V_0$: $102.5\, e_u$ (solid
line), and $105\, e_u$ (dashed line). The value of $t$ is 
selected to get
the maximum effect, $G^{\rm q}\simeq 0.24$; $t=2.731\, t_u$.
Other parameters as in Fig. 1. 

{FIG. 6.} Imaginary versus real parts of the incident contribution to
$\psi_{IT}^0(p,t)$ (empty circles), and of the transmision contribution
(filled circles), for $V_0=105\, e_u$, the value of $t$ for which the
effect is maximum ($t=2.731\, t_u$), and different values of $p$.
Other parameters as in Fig. 1. 

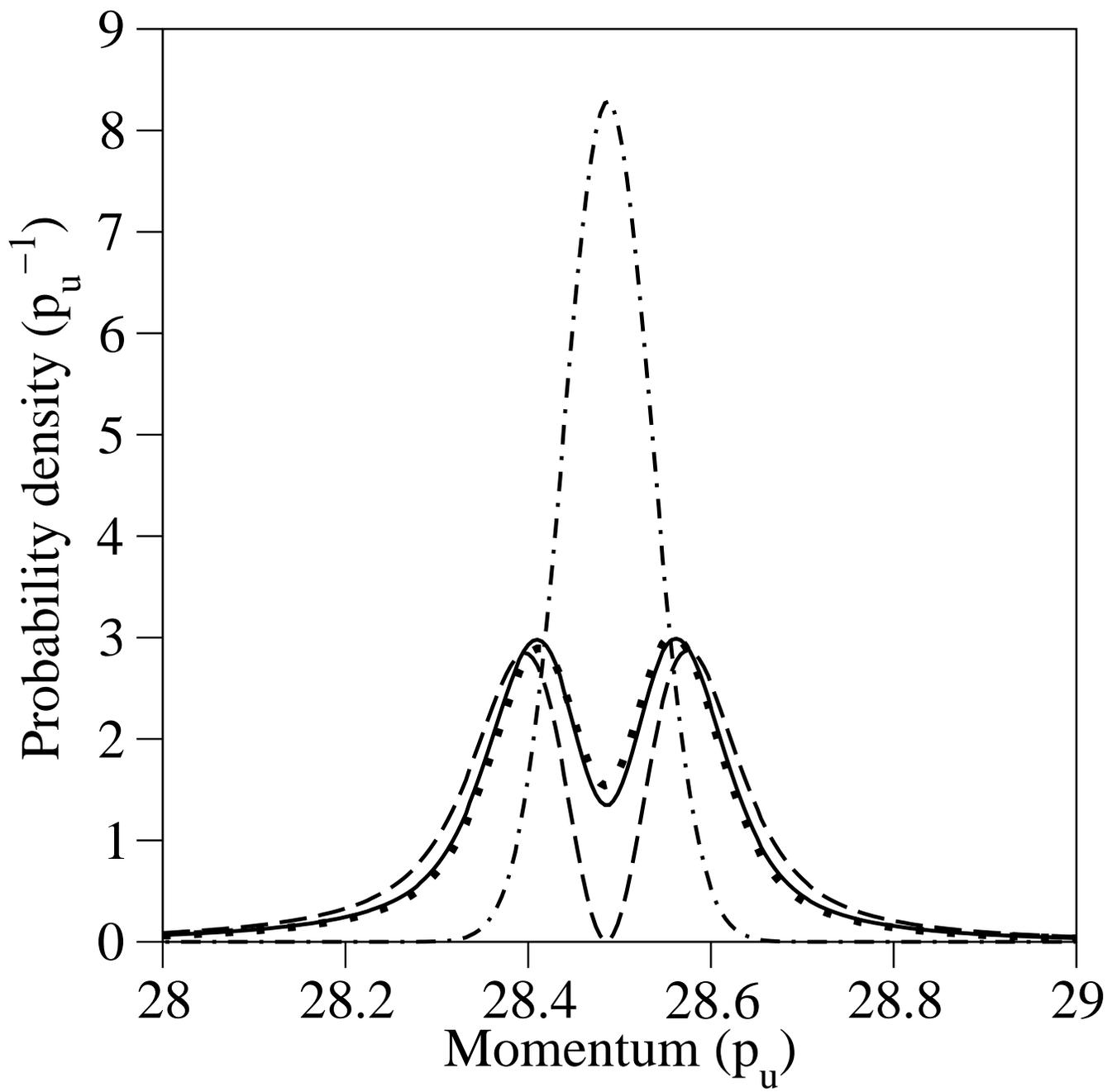
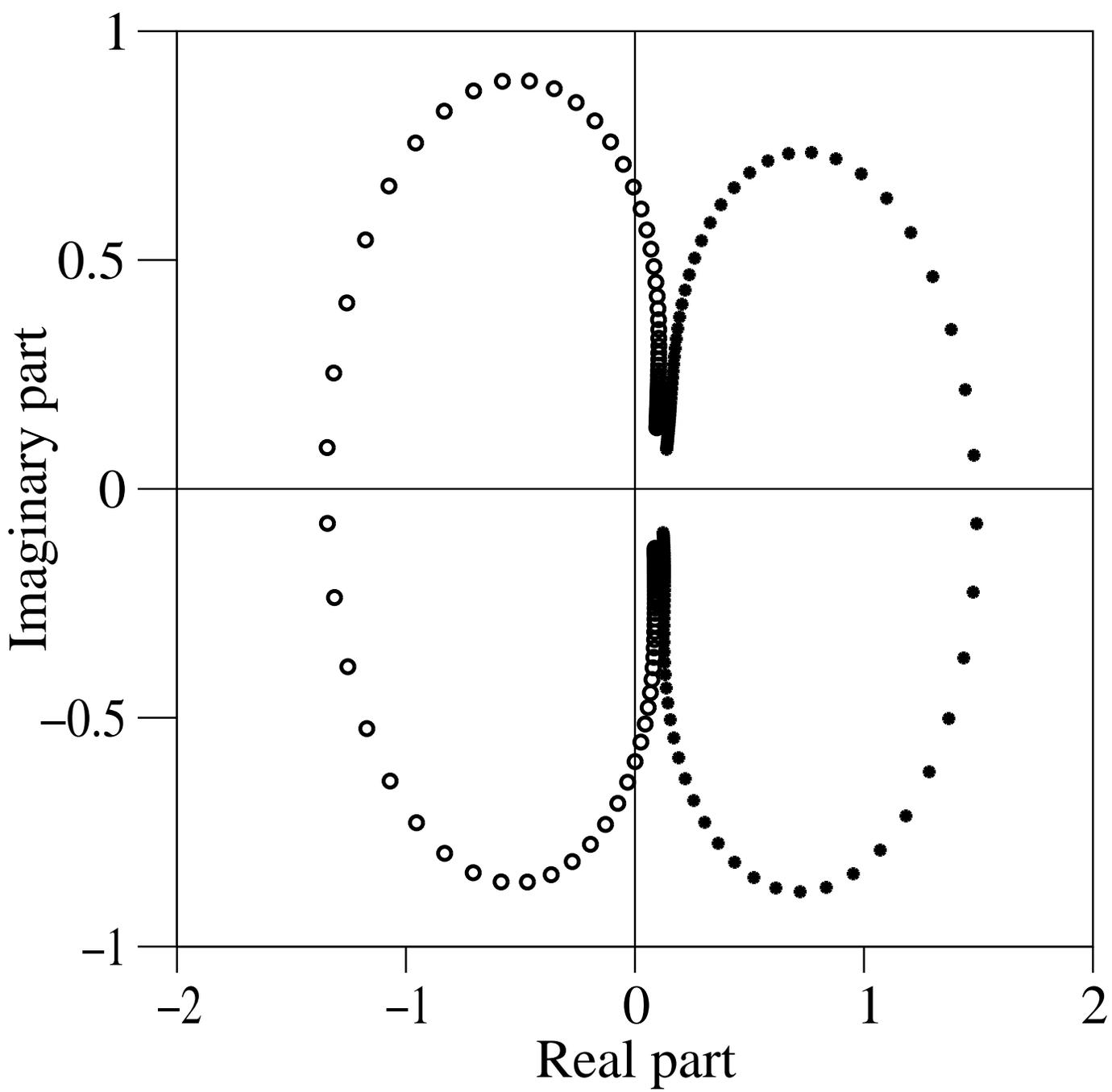
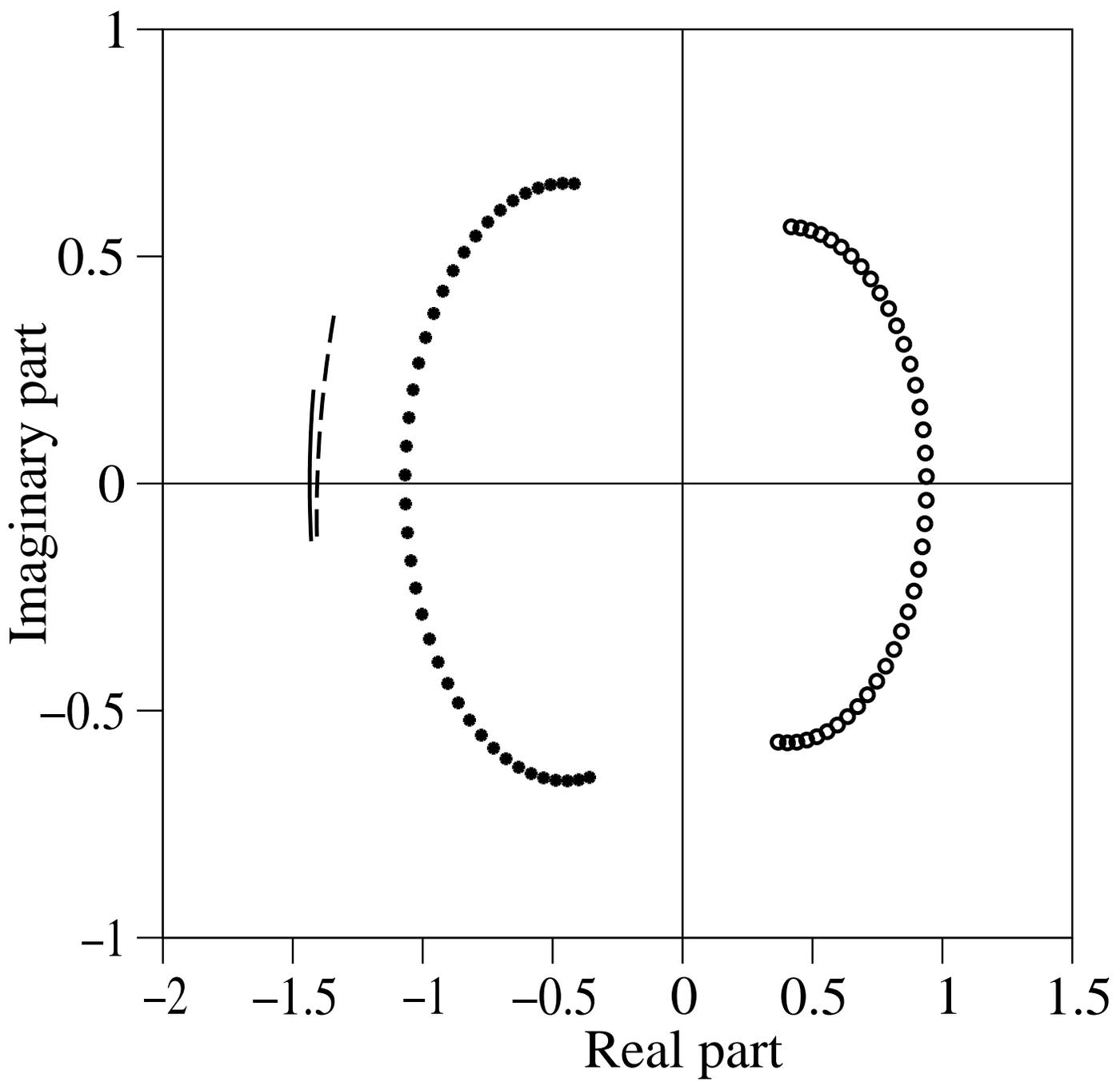
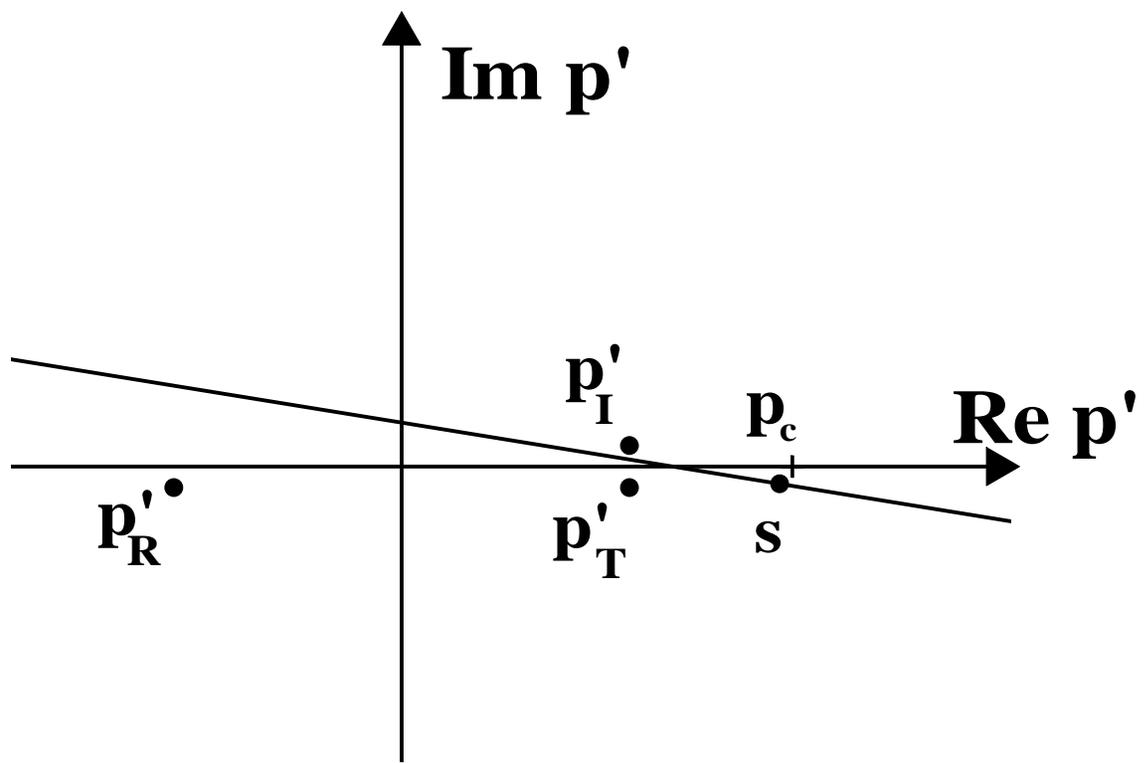
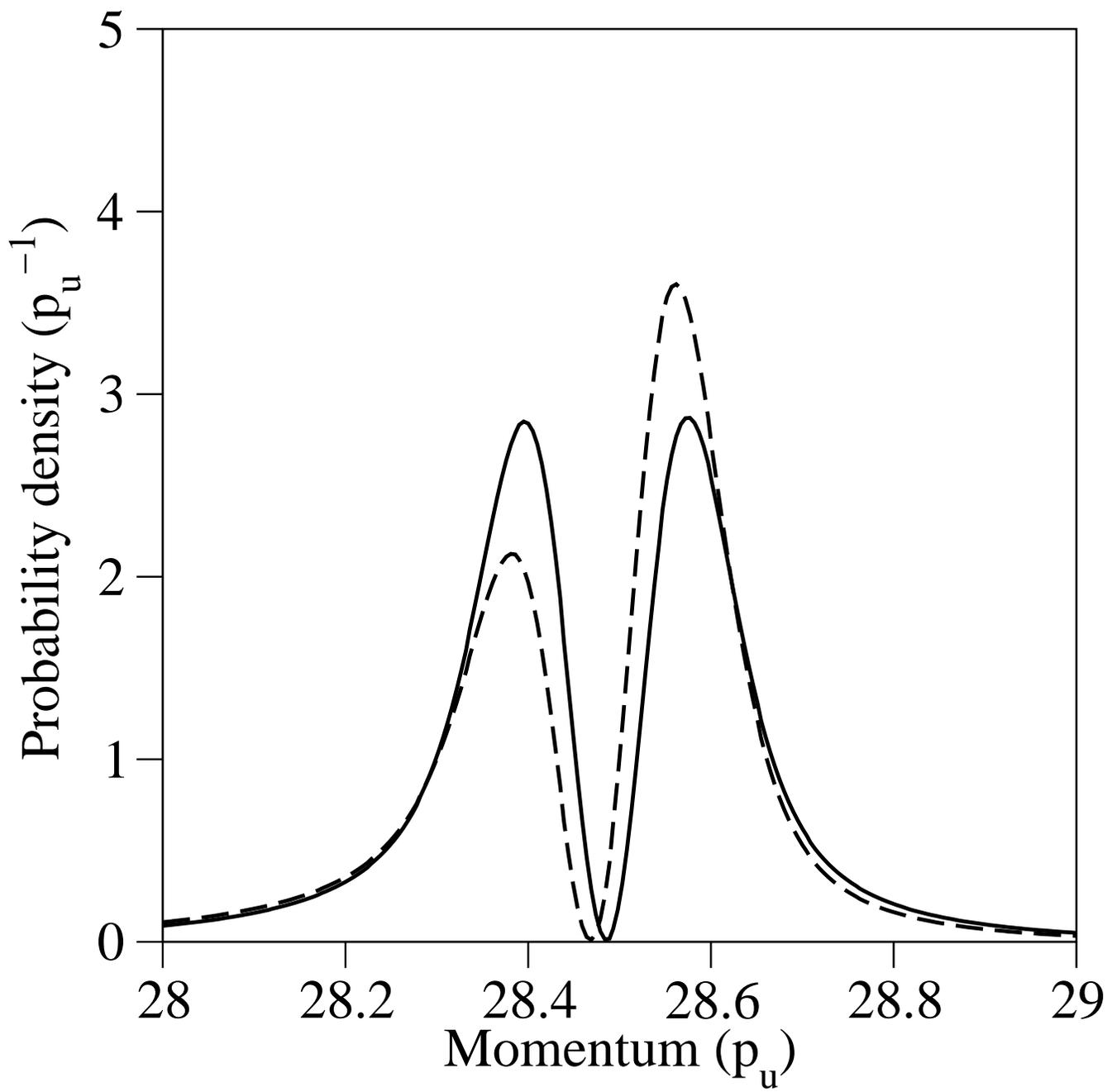
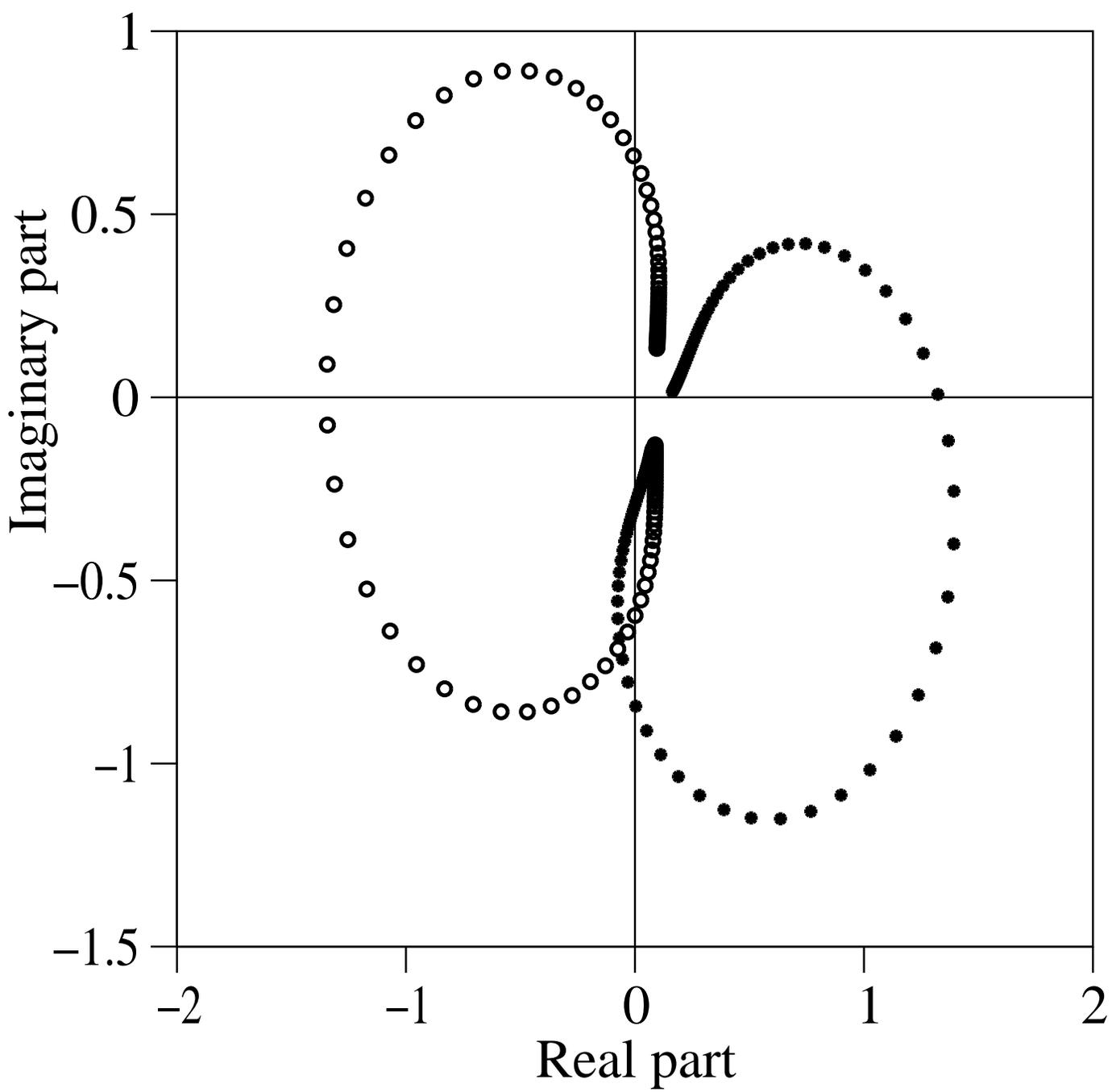
\end{document}